\documentclass[journal]{IEEEtran}
\usepackage{amsmath}
\usepackage{amssymb}
\usepackage{amsfonts}
\usepackage{bm}
\usepackage{cite}

\usepackage[acronyms,nonumberlist,nogroupskip]{glossaries}
\usepackage[table]{xcolor}
\usepackage{booktabs} 
\usepackage{makecell}
\usepackage{dirtytalk}
\usepackage{graphicx}
\usepackage{siunitx}
\usepackage[hyphens]{url}
\newglossary{symbols}{sym}{sbl}{}

\usepackage{mathtools}
\usepackage{amsthm}
\usepackage{pifont}
\ifCLASSOPTIONcompsoc
 \usepackage[caption=false,font=normalsize,labelfont=sf,textfont=sf]{subfig}
\else
 \usepackage[caption=false,font=footnotesize]{subfig}
\fi

\usepackage[hidelinks,pdfencoding=auto, psdextra]{hyperref} %
\usepackage[capitalize]{cleveref}
\DeclareSIUnit{\pu}{p.u.}
\crefname{equation}{}{}
\crefname{appendix}{Appendix}{Appendices}

\DeclareMathDelimiter{(}{\mathopen} {operators}{"28}{largesymbols}{"00}
\DeclareMathDelimiter{)}{\mathclose}{operators}{"29}{largesymbols}{"01}

\newcommand{\inbrackets}[1]{\ensuremath{\left(#1\right)}}
\newcommand{\inSet}[1]{\ensuremath{\in \left \{#1\right \}}}
\newcommand{\defineSet}[1]{\ensuremath{\left \{#1\right \}}}

\newcommand{\sizeof}[1]{\ensuremath{\left|#1\right|}}
\newcommand{\norm}[2]{\ensuremath{\left\lVert#1\right\rVert_{#2}}}
\newcommand{\twonorm}[1]{\norm{#1}{2}^2}

\newcommand{\realOf}[1]{\ensuremath{\operatorname{Re} \inbrackets{#1}}}
\newcommand{\imagOf}[1]{\ensuremath{\operatorname{Im} \inbrackets{#1}}}

\newcommand{\Rdim}[1]{\ensuremath{\mathbb{R}^{#1}}}
\newcommand{\Cdim}[1]{\ensuremath{\mathbb{C}^{#1}}}

\newcommand{\setOfBuses}{\ensuremath{\mathcal{N}}}
\newcommand{\setOfBranches}{\ensuremath{\mathcal{B}}}
\newcommand{\setOfInjector}{\ensuremath{\mathcal{I}}}

\newcommand{\setOfCycles}{\ensuremath{\mathcal{C}}}

\newcommand{\powerFlowMap}{\ensuremath{\bm{\Psi}}}
\newcommand{\powerFlowMapOf}[1]{\ensuremath{\powerFlowMap\inbrackets{#1}}}
\newcommand{\powerFlowVUMap}{\ensuremath{\bm{\Phi}}}
\newcommand{\powerFlowVUMapOf}[1]{\ensuremath{\powerFlowVUMap\inbrackets{#1}}}

\newcommand{\voltage}{\ensuremath{\bm{v}}}
\newcommand{\voltageHat}{\ensuremath{\hat{\bm{v}}}}
\newcommand{\voltageStar}{\ensuremath{\bm{v}^*}}

\newcommand{\control}{\ensuremath{\bm{u}}}
\newcommand{\controli}[1]{\ensuremath{u_{#1}}}
\newcommand{\controlis}[1]{\ensuremath{\bm{u}_{#1}}}
\newcommand{\controlopt}{\ensuremath{\bm{u}_{\text{opt}}}}
\newcommand{\controlfix}{\ensuremath{\bm{u}_{\text{fix}}}}
\newcommand{\controlstar}{\ensuremath{\bm{u}^*}}
\newcommand{\controlstarHat}{\ensuremath{\hat{\bm{u}}^*}}

\newcommand{\currenti}[1]{\ensuremath{i_{#1}}}

\newcommand{\statei}[1]{\ensuremath{\bm{x}_{#1}}}

\newcommand{\voltageMagnitudei}[1]{\ensuremath{V_{#1}}}
\newcommand{\branchAngle}{\ensuremath{\bm{\varphi}}}
\newcommand{\branchAnglei}[1]{\ensuremath{\varphi_{#1}}}

\DeclareMathOperator{\KCL}{KCL}
\DeclareMathOperator{\KVL}{KVL}

\newcommand{\KCLiOfVoltage}[1]{\ensuremath{\KCL_{#1}\inbrackets{\voltage}}}
\newcommand{\KVLiOfVoltage}[1]{\ensuremath{\KVL_{#1}\inbrackets{\voltage}}}

\newcommand{\residual}{\ensuremath{\bm{r}}}

\newcommand{\residualofVoltage}{\ensuremath{\residual \inbrackets{\voltage}}}
\newcommand{\residualofVgivenU}{\ensuremath{\residual \inbrackets{\voltage ;\control}}}

\newcommand{\residualScalar}{\ensuremath{\rho}}
\newcommand{\residualScalarofVgivenU}{\ensuremath{\residualScalar \inbrackets{\voltage ;\control}}}
\newcommand{\residualScalarofVHatgivenU}{\ensuremath{\residualScalar \inbrackets{\voltageHat ;\control}}}

\newcommand{\NNparameters}{\ensuremath{\theta}}
\newcommand{\neuralSolverMap}{\ensuremath{\hat{\powerFlowVUMap}_{\NNparameters}}}
\newcommand{\neuralSolverMapOfControl}{\ensuremath{\neuralSolverMap \inbrackets{\control}}}
\newcommand{\features}{\ensuremath{\bm{\phi}}}
\newcommand{\featuresOfControl}{\ensuremath{\features\inbrackets{\control}}}
\newcommand{\featuresLearned}{\ensuremath{\features_{\NNparameters}}}
\newcommand{\featuresLearnedOfControl}{\ensuremath{\featuresLearned\inbrackets{\control}}}

\newcommand{\sample}[2]{\ensuremath{#1^{(#2)}}}

\newcommand{\dataset}{\ensuremath{\mathcal{D}}}
\newcommand{\datasetTrain}{\ensuremath{\dataset_{\text{train}}}}

\newcommand{\loss}{\ensuremath{\mathcal{L}}}

\newacronym{PO}{PO}{Predict-then-Optimise}
\newacronym{LBFGS}{L-BFGS}{limited memory-BFGS}
\newacronym{NN}{NN}{Neural Network}
\newacronym{GNN}{GNN}{Graph Neural Network}

\newacronym{PF}{PF}{Power Flow}
\newacronym{RPF}{RPF}{Residual Power Flow}
\newacronym{OPF}{OPF}{Optimal Power Flow}
\newacronym{OC}{OC}{Operating Condition}

\newacronym{BIM}{BIM}{Bus Injection Model}

\newacronym{PINN}{PINN}{Physics-Informed Neural Network}
\newacronym{RK}{RK}{Runge-Kutta}
\newacronym{DAE}{DAE}{Differential-Algebraic Equation}
\newacronym{DE}{DE}{Differential Equation}

\newacronym{AD}{AD}{Automatic Differentiation}
\newacronym{TDS}{TDS}{Time-Domain Simulation}

\newcommand{\cmark}{\ding{51}}%
\newcommand{\xmark}{\ding{55}}%

\definecolor{BIM}{RGB}{158,202,225}
\definecolor{RPF}{RGB}{241,105,19}

\newtoggle{shortParagraph}
\toggletrue{shortParagraph}
\newtoggle{shortSection}
\toggletrue{shortSection}

\begin{document}
\bstctlcite{IEEEexample:BSTcontrol}
\title{Residual Power Flow for Neural Solvers}

\author{Jochen~Stiasny,~\IEEEmembership{Member,~IEEE,}
        and~Jochen~Cremer,~\IEEEmembership{Senior Member,~IEEE}
\thanks{J. Stiasny and J. Cremer are with the Department
of Electrical and Computer Engineering, Delft University of Technology, Delft,
Netherlands and the Austrian Institute of Technology, Vienna, Austria, e-mail: \{j.b.stiasny, j.l.cremer\}@tudelft.nl.\\
This work has been submitted to the IEEE for possible publication. Copyright may be transferred without notice, after which this version may no longer be accessible.}
}

\markboth{}%
{}

\maketitle

\begin{abstract}    
The energy transition challenges operational tasks based on simulations and optimisation. These computations need to be fast and flexible as the grid is ever-expanding, and renewables' uncertainty requires a flexible operational environment. Learned approximations, proxies or surrogates---we refer to them as Neural Solvers---excel in terms of evaluation speed, but are inflexible with respect to adjusting to changing tasks. Hence, neural solvers are usually applicable to highly specific tasks, which limits their usefulness in practice; a widely reusable, foundational neural solver is required. Therefore, this work proposes the \gls{RPF} formulation. \gls{RPF} formulates residual functions based on Kirchhoff's laws to quantify the infeasibility of an operating condition. The minimisation of the residuals determines the voltage solution; an additional slack variable is needed to achieve AC-feasibility. \gls{RPF} forms a natural, foundational subtask of tasks subject to power flow constraints. We propose to learn \gls{RPF} with neural solvers to exploit their speed. Furthermore, \gls{RPF} improves learning performance compared to common power flow formulations. To solve operational tasks, we integrate the neural solver in a \gls{PO} approach to combine speed and flexibility. The case study investigates the IEEE $9$-bus system and three tasks (AC \gls{OPF}, power-flow and quasi-steady state power flow) solved by \gls{PO}. The results demonstrate the accuracy and flexibility of learning with \gls{RPF}.
\end{abstract}


\glsresetall
\section{Introduction}

\IEEEPARstart{T}{he} energy transition changes the operation of power systems. The systems must operate at a higher throughput due to the electrification and increasing demand. At the same time, the dependency on increasingly more renewable energy leads to more uncertainty which requires grid operators to improve operational awareness and ensure security \cite{panciatici_operating_2012}. Improving situational awareness in operations requires relying more and more on repeated simulations and optimisations, which we refer to as computational \textit{tasks} \cite{konstantelos_implementation_2017,pandey_large-scale_2023}. Such a task could be, for example, to determine AC-\gls{OPF}, to perform state estimation, or to assess dynamic security. The solutions of these tasks depend on the \textit{\glspl{OC}}, such as loading conditions, and \textit{task specifications}, such as operational constraints and objectives. The dominating type of approach is numerical algorithms that are often specialised for a specific task. These algorithms exploit the problem structure, and while they have great applicability to different task specifications, \glspl{OC} are often re-solved repeatedly even if a very similar \gls{OC} had been solved before. In contrast, the recent focus on learning-based approaches, \textit{Neural Solvers}, relies on exploiting similarities between \glspl{OC} by learning shared relationships; a few of the vast number of solvers are reviewed in \cite{khaloie_review_2025, huang_applications_2023}. While neural solvers provide fast and differentiable solutions, they lack the flexibility to handle different task specifications \cite{stiasny_physics-informed_2023}.
In this work, we provide a structure to combine neural and numerical solvers for many relevant tasks in power systems to make use of each solver's strengths.

\begin{figure}
    \centering
    \includegraphics[width=0.95\linewidth]{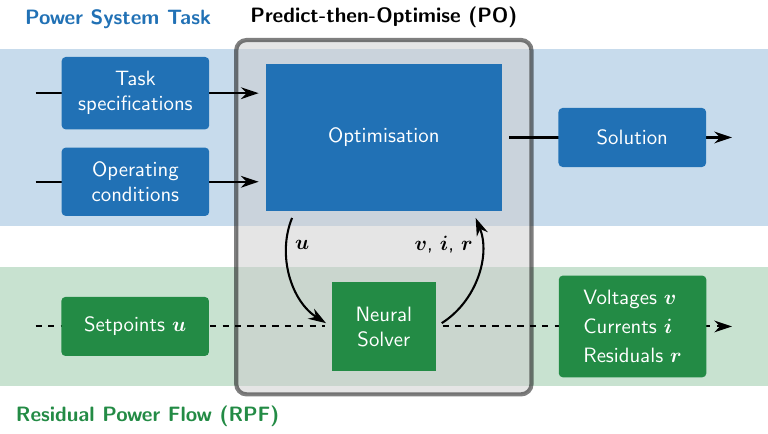}
    \caption{Proposed \gls{PO} approach: The power system task is formulated as an optimisation requiring task specifications and the operating conditions. The sub-task considers the power flow equations with the neural solver approximating the \acrfull{RPF}. 
    }
    \label{fig:task_structure}
\end{figure}

The approach to interleaving numerical and neural solvers is rooted in the following observation: Many different tasks inherit the \gls{PF} equations as constraints \cite{milano_power_2010}. Hence, satisfying the \gls{PF} constraints forms a common \textit{sub-task} that can be used under varying \glspl{OC} and task specifications. The structure of splitting a problem into a sub-task and a main task resembles decomposition approaches which have been termed feedback control or feedback optimisation in the power flow context \cite{dallanese_optimal_2016, menta_stability_2018, behr_prime_2025}. The approach of \gls{PO} \cite{elmachtoub_smart_2022} follows a similar structure; however, \gls{PO} emphasises that the sub-task is learned. We adapt this \gls{PO} approach to power system tasks as shown in \cref{fig:task_structure}: We form the sub-task of predicting the \gls{PF} solution with a neural solver. To solve the main optimisation task, we can reuse the same neural solver in a \gls{PO} fashion to \gls{PF}-related tasks such as AC-\gls{OPF} or state estimation. The so-far unanswered question is how the learning problem for solving the \gls{PF} constraint in the sub-task shall be posed to achieve high flexibility and accuracy. 



Important criteria for quality of a (learned) \gls{PF} approximation are accuracy, speed, and robustness. The latter two criteria are well covered by linearising the \gls{PF} equations \cite{molzahn_survey_2019,bolognani_fast_2015}, and data-driven methods can be used to enhance accuracy \cite{jia_overview_2023,markovic_parameterized_2023,buason_sample-based_2025}. Generally, these approximations do not yield AC-feasible \gls{PF} solutions. \glspl{NN} allow more expressive and accurate function approximations, in particular with the use of \glspl{GNN} \cite{donon_graph_2019,donon_neural_2020}, but also other advanced \gls{NN} architectures \cite{hansen_power_2023,lin_powerflownet_2024, donon_leap_2020,dogoulis_kclnet_2025}. The quest for more accurate approximations extends to providing datasets \cite{varbella_powergraph_2025}, organising competitions \cite{leyli-abadi_machine_2025}, and training foundation models \cite{hamann_foundation_2024}. Still, learned solution will retain errors which causes AC-infeasibility of the \gls{PF}. While these errors could be bounded \cite{venzke_learning_2020}, and reduced with special training design \cite{nellikkath_physics-informed_2022}, the recovery of AC-feasible \gls{PF} solutions requires additional computations \cite{baker_emulating_2022,taheri_ac_2024}. The 
 \gls{PO} approach enables a different path to handle prediction errors and AC-feasibility: We designate that the main optimisation task needs to address the feasibility of the \gls{PF}. Meanwhile, the neural solver aims at predicting \gls{PF} solutions that are \say{close to feasible} with desirable prediction error characteristics.


This proposition of focusing on infeasible \gls{PF} solutions requires a continuous quantification of infeasibility instead of a binary quantification (feasibility versus infeasibility). To this end, the contributions of this paper are:
\begin{itemize}
    \item \gls{RPF} formulation defining a residual function to quantify power flow infeasibility. 
    \item \acrfull{PO} approach with neural solvers for \gls{RPF} as a foundation for a variety of power system tasks constrained by \gls{PF}. 
    \item Analysis of the \gls{RPF} formulation's impact on the approximation quality of neural solvers.
\end{itemize}
To demonstrate the proposed approach, we learn the \gls{RPF} solution with a neural solver across a range of \glspl{OC} for the IEEE $9$-bus system. We then apply the neural solver to several power system tasks (AC-feasible \gls{PF}, quasi-steady-state \gls{PF}, AC-\gls{OPF}) to illustrate the flexibility of the \gls{PO} approach.

The paper is structured as follows: \Cref{sec:formulation} presents the \gls{RPF} formulation. The neural solver and \gls{PO} approach are defined in \cref{sec:neural_solver,sec:PO} and tested in the case study in \cref{sec:learning_problem,sec:experiments} respectively. We discuss the consequences of adopting \gls{RPF} in \cref{sec:discussion} and conclude in \cref{sec:conclusion}.

\section{Residual Power Flow (RPF) Formulation}\label{sec:formulation}

This section introduces the proposed \gls{RPF} formulation. We begin by defining infeasibility and a brief exposition of Kirchhoff's laws, followed by introducing the variables and constraints in \gls{RPF}. Subsequently, we describe the modelling of the current injectors and the definition of the \gls{RPF} solution.

\subsection{Feasibility of AC-PF and the role of bus types}
A feasible \gls{PF} is defined by the combination of voltages \voltage{} and controls \control{} that satisfy a set of equality constraints \powerFlowMap{}
\begin{align}\label{eq:feasible_PF}
    \powerFlowMapOf{\voltage, \control} = \bm{0}.
\end{align}
The controls \control{} include setpoints of generation and load and thereby determine the power (or current) injections for an \gls{OC}. If $\powerFlowMapOf{\voltage, \control} \neq \bm{0}$, we consider the \gls{PF} infeasible. While the underlying physics, namely Kirchhoff's laws, universally prescribe the constraints in \powerFlowMapOf{\voltage{}, \control{}}, there are different mathematical representations possible as reviewed in \cite{molzahn_survey_2019}. The most common representation is the \gls{BIM} with different variable formulations, followed by the \say{DistFlow} \cite{baran_optimal_1989,baran_optimal_1989-1} or \say{Branch flow} \cite{farivar_branch_2013} model. \textit{Solving
\gls{PF}} implicitly means finding a feasible solution to \powerFlowMapOf{\voltage, \control}. To do so, modellers make an additional central representation choice: Buses in a grid have types---a slack bus, PV and PQ buses, and a reference bus\footnote{The reference bus is only necessary in the \gls{BIM} and is often selected to coincide with the slack bus.}. The bus types define the known and unknown variables, as shown in \cref{tab:bus_types}, and ensure a fully determined system of equations with equal number of constraints and variables. 
Introducing bus types effectively reduces the number of constraints in \cref{eq:feasible_PF} by selecting the control variables $Q$ at PV-buses and $P$ and $Q$ at the slack bus after solving for the voltage variables. Thereby, the affected constraints are always satisfied.
\renewcommand{\arraystretch}{1.5}
\begin{table}[ht]
    \caption{Known (\cmark), unknown (\xmark) and derived (-) variables for different bus types in a bus injection formulation.}
    \centering
    \begin{tabular}{ccccc}
        \toprule
        Bus type & \thead{voltage\\magnitude $V$} &  \thead{voltage\\angle $\theta$}  & \thead{active\\power $P$} & \thead{reactive\\power $Q$}\\
        \midrule
        PQ & \xmark & \xmark & \cmark & \cmark \\
        PV & \cmark & \xmark & \cmark & - \\
        slack & \cmark & \cmark & - & -\\
        \bottomrule
    \end{tabular}
    \label{tab:bus_types}
\end{table}
\renewcommand{\arraystretch}{1.2}

When we learn a solution $\voltageHat{}\inbrackets{\control{}}$, we will encounter constraint violations as the approximation \voltageHat{} will carry some error $|\voltageHat{} - \voltage{}|$. Because of the bus type assignment, these constraint violations can be compensated at PV and slack buses by adjusting the control variables $\hat{\control{}}\inbrackets{\voltageHat{}}$ accordingly. However, the adjusted controls $\hat{\control{}}$ then also differ from the exact, feasible solution \control{}. As a result, a learned approximation using the \gls{BIM} with bus types will lead to some satisfied constraints and correct voltages, but also unsatisfied constraints, erroneous voltages and wrongly adjusted control values. In summary, a highly \textit{asymmetric} distribution of the approximation errors and no clear notion of the infeasibility of the solution. 

For numerical methods, the asymmetry, which the bus types cause, is less critical as the system of equations is solved to very high accuracy. Still, the assignment of a slack bus \cite{dhople_reexamining_2020,milano_dynamic_2024} or reactive power limits that lead to switching between PV and PQ buses \cite{zeng_accuracy_2023,milano_power_2010} can cause ambiguities and difficulties when solving \gls{PF} in this representation. For learning problems, however, steps that affect the representation become critical. Such steps can be feature engineering, the definition of learning objectives, and the choice of model architectures \cite{bengio_representation_2013}. Ideally, the representation captures the geometry and symmetries of the underlying problem \cite{hamilton_graph_2020, testa_geometric_2025}. Disregarding these representational choices can require more data in the training, perform worse in unseen scenarios, or lead to decision that are clearly contradicting the underlying physics.

Therefore, we have two objectives in the development of the \gls{PF} formulation: First, to define a clear notion of infeasibility and second, to avoid the assignment of bus types which causes representational asymmetries. By fulfilling these objectives, we can define a meaningful value to quantify the infeasibility of a \gls{PF} and obtain a well-defined map $\control{} \mapsto \voltage{}$ that we will subsequently learn with a neural solver.

\subsection{Kirchhoff laws}
\begin{figure}[ht]
    \centering
    \includegraphics[width=0.8\linewidth]{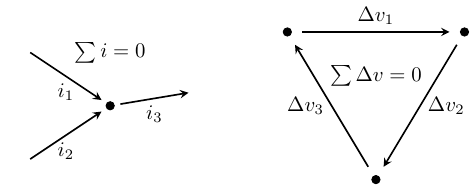}
    \caption{Kirchhoff's current and voltage law.}
    \label{fig:kirchhoffs_laws}
\end{figure}

The governing laws that need to hold in an electric circuit are Kirchhoff's current and voltage law, schematically shown in \cref{fig:kirchhoffs_laws}.
The current law states that at a node $j$ the current injections $i_k$ from the set of connected injectors $\setOfInjector{}_j$ sum to 0
\begin{subequations}\label{eq:Kirchhoffs_laws}
\begin{align}\label{eq:KCL_gerenal}
    \sum_{k \in \setOfInjector{}_j} i_k &= 0.
\end{align}
Kirchhoff's voltage law demands that the potential differences, expressed in terms of the voltage difference $\Delta v_k$ for line $k$, around a cycle sum to 0 
\begin{align}\label{eq:KVL_gerenal}
    \sum_{k \in \setOfCycles{}_j} \Delta v_k &= 0.
\end{align}
The lines are part of the $j$-th cycle $\setOfCycles{}_j$ in the electric circuit. Based on these two laws, the relation between currents and voltages in the circuit can be determined.
\end{subequations}

\subsection{Variables and Residual Functions in RPF}
We apply Kirchhoff's laws \cref{eq:Kirchhoffs_laws} to the power grid setting to derive the set of variables and constraints of the proposed \gls{RPF} formulation\footnote{We use the phasor representation of a balanced network; an imbalanced representation could be formulated by adding variables and constraints for each phase.}. We begin by forming the vector of voltage variables \voltage{}. It combines the voltage magnitudes $\voltageMagnitudei{k}\in \Rdim{}_{>0}$ of each bus $k$ in the set of buses \setOfBuses{} and the \textit{branch angle} $\branchAnglei{k} \in \left(-\frac{\pi}{2},\frac{\pi}{2}\right) \subset \Rdim{}$, that is the angle difference over branch $k$, in the set of branches \setOfBranches{}. As the branches represent undirected edges in the graph, the sign of $\branchAnglei{k}$ depends on the assigned direction between the buses $i$ and $j$ as $\branchAnglei{i\rightarrow j} = - \branchAnglei{j \rightarrow i}$, but this decision has no further effects. This construction yields the vector $\voltage{} \in \Rdim{\sizeof{\setOfBuses{}}}_{>0} \times \left(-\frac{\pi}{2},\frac{\pi}{2}\right)^{\sizeof{\setOfBranches{}}}$
\begin{align}
    \voltage{} &= \begin{bmatrix}
        \voltageMagnitudei{1} & \hdots& \voltageMagnitudei{\sizeof{\setOfBuses{}}} & \branchAnglei{1} & \hdots & \branchAnglei{\sizeof{\setOfBranches{}}}
    \end{bmatrix}^\top.
\end{align}

We formulate a set of residual functions $\residual{}\inbrackets{\voltage{}, \control{}}$ arising from Kirchhoff's laws \cref{eq:Kirchhoffs_laws} that correspond to \powerFlowMapOf{\voltage{}, \control{}} in \cref{eq:feasible_PF}. For the current law \cref{eq:KCL_gerenal}, we define the complex residual function $\KCLiOfVoltage{n} \in \Cdim{}$ for each bus
\begin{subequations}\label{eq:KHLs_new}
\begin{align}\label{eq:KCL_new}
    \KCLiOfVoltage{n}&:=\, \sum_{k \in \setOfInjector{}_{n}} \currenti{k} \inbrackets{\voltage{} ; \controlis{k}, \statei{k}}, \; \forall n \inSet{1, \hdots, \sizeof{\setOfBuses{}}}
\end{align}
where $\currenti{k} \inbrackets{\voltage{} ; \controlis{k}, \statei{k}} \in \Cdim{}$ represents the current injection of each component connected to bus $n$. The control variables \controlis{k} and state variables \statei{k} can be present in the current calculation depending on the component type, as we will detail in \cref{subsec:current_injectors}. Similarly, we formulate a residual function $\KVLiOfVoltage{\ell} \in \Rdim{}$ for each cycle in the network
\begin{align}\label{eq:KVL_new}
    \KVLiOfVoltage{\ell}&:=\, y_\ell \sum_{k \in \setOfCycles{}_{\ell}} \branchAnglei{k}\inbrackets{\voltage{}}, \quad \forall \ell \inSet{1, \hdots, \sizeof{\setOfCycles{}}}.
\end{align}
\end{subequations}
The residual yields a scalar instead of a complex number, as the voltage magnitude is matched by construction in a cycle. We scale the angle mismatch by a factor $y_\ell$
\begin{align}
    y_\ell = \imagOf{\frac{1}{\sum_{k \in \setOfCycles{}_{\ell}} r_k + j x_k }}
\end{align}
which represents the imaginary part of the combined admittance of the branches in the cycle. Thereby, \KVLiOfVoltage{\ell} represents a current that can be compared to the current residuals \KCLiOfVoltage{n}. In contrast to \cref{eq:KCL_new}, we omitted dependencies of \branchAnglei{k} from control or state variables, but these could easily be added if branches are used that can control the branch angle.

Based on \cref{eq:KHLs_new}, we form the residual function \residualofVoltage{}
\begin{align}\label{eq:residual_vector}
    \residualofVoltage{}&:= \begin{bmatrix}
        \realOf{\KCLiOfVoltage{1}}\\
        \vdots\\
        \realOf{\KCLiOfVoltage{\sizeof{\setOfBuses{}}}}\\
        \imagOf{\KCLiOfVoltage{1}}\\
        \vdots \\ \imagOf{\KCLiOfVoltage{\sizeof{\setOfBuses{}}}}\\ \KVLiOfVoltage{1} \\ \vdots \\ \KVLiOfVoltage{\sizeof{\setOfCycles{}}}
    \end{bmatrix}, \quad \residual{} \in \Rdim{2\sizeof{\setOfBuses{}} + \sizeof{\setOfCycles{}}}.
\end{align}
For any power flow state $\inbrackets{\voltage{}, \control{}}$ to be feasible, we will require $\residualofVgivenU{} = \bm{0}$ to satisfy Kirchhoff's laws. Note that so far, there has been no need to define bus types. The use of branch angles instead of voltage angles at the buses eliminates the need to designate a reference bus angle. However, as the branch angles do not necessarily fulfil Kirchhoff's voltage law \cref{eq:KVL_gerenal}, we need the additional constraints in \cref{eq:KVL_new} compared to the \gls{BIM}; the \gls{RPF} formulation is similar to the branch flow model in this regard. The following subsection describes the modelling of current injections \currenti{k} by which we avoid the distinction between slack, PV or PQ buses. 

\subsection{Unified modelling of current injectors}\label{subsec:current_injectors}
The following describes how we model all components such as generators, loads, lines, and transformers as generic current injectors that define their current injection $\currenti{k}\inbrackets{\voltage{}; \controlis{k}}$ for a given setpoint \controlis{k} only in dependence of \voltage{}. Thereby, we avoid the need to make distinctions of bus types in \cref{eq:residual_vector}.

In power system dynamics, the modelling of current injectors is often divided into two classes: static injectors such as lines and loads, and dynamic injectors such as generators and inverters \cite{sauer_power_1998}. The distinction arises from the presence of dynamic states \statei{k} for a component $k$ which affect the algebraic relationship $h_k$ for the current injection $\currenti{k}\in \Cdim{}$
\begin{subequations}
    \begin{align}
        \text{Static:} \quad \currenti{k} &= h_k\inbrackets{\voltage{}, \controlis{k}}\label{eq:static_injector} \\
        \text{Dynamic:} \quad \currenti{k} &= h_k\inbrackets{\voltage{}, \controlis{k}, \statei{k}}.\label{eq:dynamic_injector}
    \end{align}
\end{subequations}
Besides, both injector types depend on the voltages \voltage{} in the network, however, usually only the variables related to the component's terminal bus(es) are relevant. All setpoints to control the component are collected in \controlis{k}.

The following uses the view, that \textit{the \gls{PF} solution} is equivalent to the steady state solution of the differential equations that govern the dynamics of the system and components
\begin{align}\label{eq:steady_state_component}
    \frac{d}{dt}\bm{x}_k = \bm{f}_k(\bm{v}, \controlis{k}, \bm{x}_k).
\end{align}
Therefore, we require for feasible \glspl{PF} $\frac{d}{dt}\bm{x}_k = 0$. This steady-state solution\footnote{It is not guaranteed that a steady-state solution exists or is unique for any given \voltage{} and \controlis{k}. We will not cover such cases in this work as performing a steady-state power flow calculation becomes questionable in itself. Any software implementations should consider though how to handle such cases.} $\statei{k}^{SS}(\voltage{}, \controlis{k})$ can be found by solving the root-finding problem $\bm{f}_k = \bm{0}$. We substitute $\statei{k}$ by $\statei{k}^{SS}$ in \cref{eq:dynamic_injector}
\begin{align}\label{eq:dynamic_injector_SS}
    \currenti{k}^{SS} &= h_k\inbrackets{\voltage{}, \controlis{k}, \statei{k}^{SS}\inbrackets{\voltage{}, \controlis{k}}}.
\end{align}
Thereby, \cref{eq:dynamic_injector_SS} becomes structurally equivalent to the static injector in \cref{eq:static_injector} as only \voltage{} and \controlis{k} are required to compute the current injection. Essentially, we treat all dynamic components as being in the steady state, and hence, as static.

From a physical perspective, PV-buses assume generators to be perfect voltage sources whereas we model them as current sources with a strong voltage dependency. As a consequence, the voltage magnitude at the terminal \voltageMagnitudei{k} is not fixed any more. Instead, we fix a voltage reference as part of the control variable \controlis{k}. Hence, the presented formulation of the current injectors eliminates the need to distinguish between PV and PQ buses and aligns the entire \gls{PF} formulation closer with the dynamic modelling of power systems.

When solving \gls{RPF}, we require computing the partial derivatives $\partial \currenti{k} / \partial \voltage{}$ and $\partial \currenti{k} / \partial \controlis{k}$. The calculation is straightforward for algebraic expression of $h_k\inbrackets{\voltage{}, \controlis{k}}$. When the problem $\bm{f}_k = \bm{0}$ does not have an analytical solution, we need to revert to differentiable root-finding solvers or explicit approximations when computing $\statei{k}^{SS}$ or $\currenti{k}^{SS}$. For fast and simple implementations, it will often be preferable to use explicit approximations, for example, by using polynomials or \glspl{NN} as a function of \voltage{} and \controlis{k}. The use of full dynamic models is also possible.

\subsection{Solving RPF by minimising the residual norm \texorpdfstring{$\residualScalar{}$}{\textrho}}\label{subsec:residual_minimisation}
To define the solution of the \gls{RPF}, we collect all control variables \controlis{k} of the $K$ components connected to the grid in the control vector $\control{}\in \Rdim{m}$
\begin{align}
    \control{} = \begin{bmatrix}
        \controlis{1}^{\top} & \hdots & \controlis{K}^{\top} 
    \end{bmatrix}^{\top}
\end{align}
Given the setpoints \control{}, we can evaluate the residual \residualofVgivenU{} as well as its norm $\residualScalarofVgivenU  \in \Rdim{}_{\geq 0}$
\begin{align}\label{eq:non-linear_least_squares}
    \residualScalarofVgivenU  := \frac{1}{2} \norm{\residualofVgivenU}{W_r}^2 = \frac{1}{2} \residualofVgivenU{}^\top \, W_r \residualofVgivenU{}.
\end{align}
A positive-definite weighting matrix $W_r$ can be used to control the relative importance of the residual terms; we set $W_r$ equal to the identity matrix. 
The scalar \residualScalarofVgivenU{} provides a metric of infeasibility and feasibility implies $\residualofVgivenU{} = \bm{0}$. We define the solution of the \gls{RPF} \voltageStar{} as a voltage \voltage{} that leads to the \textit{least infeasible} \gls{PF}
\begin{align}
    \voltageStar := \arg \min_{\voltage{}} &\; \residualScalarofVgivenU.\label{eq:rpf_unconstrained}
\end{align}
This formulation implies that for any given \control{}, it cannot be guaranteed that there exists a set of voltages \voltage{} that yields a feasible \gls{PF}. Instead, we find an \gls{RPF} solution \voltageStar{} that minimises the notion of infeasibility quantified by \residualScalarofVgivenU{}.


The solution of the \gls{RPF} is AC-feasible if and only if all constraints from Kirchhoff's law are satisfied, that is $\residualScalar{} = 0$. To find an AC-feasible \gls{PF} solution, we need to adjust a slack variable $u_s$ out of the control variables \control{} such that $\residualScalar{} = 0$, which leads to
\begin{align} \label{eq:power_flow_solution}
    \min_{u_s} \left(  \min_{\voltage{}} \; \residualScalarofVgivenU \right).
\end{align}

The formulation \cref{eq:power_flow_solution} shows that we can avoid the need for a slack bus as introducing a slack variable is sufficient. This subtle difference between a slack \textit{bus} and \textit{variable} was pointed out in \cite{milano_power_2010} but is crucial to achieve the desired \gls{PF} formulation without any bus types. Moreover, this understanding aligns well with the power system dynamics view, in which an arbitrary control input \control{} usually does not yield a steady-state solution at nominal frequency $\omega_0$. Instead, the frequency $\omega$ will deviate from $\omega_0$ to balance the system as some current injections of the components have a dependency on the system frequency. The system frequency $\omega$ becomes part of \control{} and acts naturally as the slack variable $u_s$. 
\section{Neural Solvers for Residual Power Flow}\label{sec:neural_solver}

The following provides the conceptual setting of learning a neural solver for \gls{RPF}, while \cref{sec:PO} describes the integration of the neural solver into the \gls{PO} approach.

The map \powerFlowVUMap{} relates \control{} and \voltageStar{} and is defined by the solution to \cref{eq:rpf_unconstrained}
\begin{multline}
    \powerFlowVUMap : \Rdim{m} \mapsto \Rdim{\sizeof{\setOfBuses{}}}_{>0} \times \left(-\frac{\pi}{2},\frac{\pi}{2}\right)^{\sizeof{\setOfBranches{}}},\\ \control \mapsto \voltageStar = \arg \min_{\voltage{}}\residualScalarofVgivenU{}. 
\end{multline}
We denote the approximation to \powerFlowVUMap{} by a neural solver as \neuralSolverMap{}.

\subsection{Functional form of Neural Solvers}
The neural solver \neuralSolverMap{} can take many functional forms, a generic form follows   
\begin{align}
    \neuralSolverMapOfControl = A_{\NNparameters} \; \featuresLearnedOfControl \label{eq:neuralSolverFunctionalForm}
\end{align}
where $A_{\NNparameters} \in \Rdim{\inbrackets{\sizeof{\setOfBuses{}} + \sizeof{\setOfBranches{}}} \times F}$ represents a learnable matrix that linearly transforms a set of features $\featuresLearnedOfControl{} \in \Rdim{F}$ into the voltage prediction.


The simplest feature construction is linear in \control{}
\begin{align}\label{eq:neural_solver_linear}
    \featuresOfControl &= \begin{bmatrix}
        1 & \controli{1} & \cdots & \controli{m} 
    \end{bmatrix}
\end{align}
while a more powerful representation can be achieved with learned features, for example, from a feed-forward \gls{NN}
\begin{align}\label{eq:neural_solver_learned}
    \featuresLearnedOfControl &= l_{\NNparameters}^{(L)}\circ \cdots \circ l_{\NNparameters}^{(1)}\inbrackets{\control}
    \intertext{where the $L$ layers}
    l_{\NNparameters}^{(k)}\inbrackets{\bm{z}} &= \sigma \inbrackets{W^{(k)} \bm{z} + \bm{b}^{(k)}} 
\end{align}
combine a linear transformation with an element-wise non-linear function $\sigma$. The weights $W^{(k)}$ and biases $\bm{b}^{(k)}$ form the learnable parameters $\NNparameters = [W^{(k)}, \bm{b}^{(k)}]_{1\leq k \leq L}$. 

\subsection{Learning setting for Neural Solvers}
To train the selected neural solver architecture, we provide a training dataset \datasetTrain{} with $\sizeof{\datasetTrain{}}$ pairs of \control{} and \voltageStar{}
\begin{align}
    \datasetTrain = \defineSet{ \inbrackets{\sample{\control{}}{j}; \voltage{}^{*,(j)} } }_{1\leq j \leq \sizeof{\datasetTrain{}}}.
\end{align}
The dataset can contain pairs that have a AC-feasible power flow solutions, that is $\residualScalar{}\inbrackets{\voltage{}^{*,(j)}, \sample{\control{}}{j}} = 0$. However, pairs of \control{} and \voltageStar{} with non-zero residuals are also permissible. Such pairs can be used to improve the robustness of the approximation with respect to control inputs that are not AC-feasible as we demonstrate in \cref{subsec:learning_infeasible_OCs}.

To fit the learnable parameters, we form the loss \loss{}
\begin{align}\label{eq:training_function}
    \loss{}\inbrackets{\NNparameters; \datasetTrain{}} = \frac{1}{\sizeof{\datasetTrain}} \sum_{j=1}^{\sizeof{\datasetTrain}} \twonorm{\neuralSolverMap{}\inbrackets{ \sample{\control}{j}} - \voltage{}^{*,(j)}}
\end{align}
and the parameters are found by solving
\begin{align}\label{eq:loss_minimisation}
    \NNparameters^* = \arg \min_{\NNparameters} \loss{}\inbrackets{\NNparameters; \datasetTrain{}}.
\end{align}
For non-learned features, \cref{eq:loss_minimisation} is a least-squares problem that can be solved directly. For learned features gradient descent methods such as Adam \cite{kingma_adam_2015} or the \glsentrylong{LBFGS} algorithm \cite{liu_limited_1989} can be applied to optimise \cref{eq:loss_minimisation}. Extensions of the loss function as in physics-informed \glspl{NN} \cite{raissi_physics-informed_2018} are possible. Standard learning procedures of validation and testing with corresponding datasets should be used. 
\section{Predict-then-optimise with RPF}\label{sec:PO}

With the definition of the neural solver \neuralSolverMap{} as above, it is straightforward to formulate the \gls{PO} problem in a form that is applicable to a wide range of power system tasks.
\begin{subequations}\label{eq:PO_formulation_generic}
\begin{alignat}{2}
\min_{\controlopt} \quad & f(\control)  + \lambda \residualScalar{}\inbrackets{\voltageHat, \control}&&\\
\text{s.t.}\quad & \voltageHat =\neuralSolverMapOfControl&&\\
& \bm{g}\inbrackets{\voltageHat, \control{}} \leq \bm{0}&&
\end{alignat}
\end{subequations}
The objective function consists of a cost function $f\inbrackets{\control}$ and the \gls{PF} residual $\residualScalar{}\inbrackets{\voltageHat, \control}$ weighted by $\lambda$. Other constraints can be included in $\bm{g}\inbrackets{\voltageHat, \control{}}$. The decision variables \controlopt{} will often be a subset of all available control variables \control{}, for example, load setpoints might not be adjusted unless load shedding is considered. Hence, we distinguish between decision variables \controlopt{} and non-decision variables \controlfix{}
\begin{align}
    \control{} =\begin{bmatrix}
    \controlopt{}^\top & \controlfix{}^\top
\end{bmatrix}^\top.
\end{align}
Based on \cref{eq:PO_formulation_generic}, we will subsequently formulate a number of power system calculation tasks. The resulting optimisation problems can be solved with an optimiser of one's choice as the gradient and Hessian of \residualScalar{} can be computed using automatic differentiation of the neural solver. 

\subsection{PO for AC-feasible Power Flow}
As elaborated in \cref{subsec:residual_minimisation}, finding a feasible power flow corresponds to adjusting a slack variable $\controlopt = \controli{s}$ to yield $\residualScalar{}=0$, which simplifies \cref{eq:PO_formulation_generic} to
\begin{subequations}\label{eq:PO_power_flow}
\begin{alignat}{2}
\min_{u_s} \quad & \residualScalar{}\inbrackets{\voltageHat, \control{}} &&\\
\text{s.t.}\quad & \voltageHat =\neuralSolverMapOfControl.
\end{alignat}
\end{subequations}
The choice of the slack variable is free, it can be a single variable or a combination of a variables. The later allows to easily formulate a distributed slack as suggested in \cite{dhople_reexamining_2020}.

\subsection{Quasi-steady state approximations}
In the study of long-term dynamics, the control variables $\control{}$ can deviate from their initial setpoint $\control{}_0$ depending on the system frequency $\omega$. Typical examples are the frequency control of generators and frequency-dependent loads. The assumption for the calculation is that the faster dynamics have settled at a steady state, see \cite{cutsem_voltage_1998} for an introduction. We can express such setting by using $\omega$ as the slack variable and adding the frequency-dependent control policy $\bm{\pi}$ to adjust \control{} 
\begin{subequations}\label{eq:PO_qss}
\begin{alignat}{2}
\min_{\omega} \quad & \residualScalar{}\inbrackets{\voltageHat, \control{}} &&\\
\text{s.t.}\quad & \voltageHat =\neuralSolverMapOfControl\\
& \control{} = \bm{\pi}\inbrackets{\control{}_0, \omega, \voltageHat}.
\end{alignat}
\end{subequations}

\subsection{AC-Optimal Power Flow}
An advanced use of \gls{PO} could be the solution of an AC-\gls{OPF} problem with a quadratic cost function $f\inbrackets{\control{}}$ parametrised by $Q$ and $\bm{q}$ and a set of inequality constraints $\bm{g} \inbrackets{\voltageHat, \control{}}$ which include operational constraints  on setpoints $\bm{g}_u \inbrackets{\control{}}$, voltage magnitudes and branch angles $\bm{g}_v \inbrackets{\voltageHat}$, and current limits $\bm{g}_i \inbrackets{\voltageHat, \control{}}$
\begin{subequations}\label{eq:PO_AC_OPF}
\begin{alignat}{2}
\min_{\controlopt} \quad & \control{}^\top Q \control{} + \bm{q}^\top \control{}+ \lambda \residualScalar{}\inbrackets{\voltageHat, \control{}} &&\\
\text{s.t.}\quad & \voltageHat =\neuralSolverMapOfControl\\
&  \bm{g}\inbrackets{\voltageHat, \control{}} \leq \bm{0}.
\end{alignat}
\end{subequations}
Limits on the voltage magnitude and branch angles will be evaluated based on the neural solver's approximation \voltageHat{} and the current limits by first calculating $\currenti{k} = h_k\inbrackets{ \voltageHat{}, \controlis{k}}$ for the relevant components. The weighting factor $\lambda$ controls the impact of the \gls{PF} residual, a high value penalises \gls{PF} constraint violations more.

\section{Case study: Learning RPF with Neural Solvers}\label{sec:learning_problem}

In this case study we demonstrate the implication of switching from a \gls{BIM} to \gls{RPF} when learning \glspl{PF} with neural solvers. We analyse the distribution of approximation errors across the voltage variables \voltage{} and across the residuals \residual{}. These analyses highlight the effect of asymmetries in the \gls{PF} formulation and the treatment of infeasibility which \gls{RPF} improves.

\subsection{Experiment setup}
All experiments are implemented in Julia \cite{bezanson_julia_2017} and run on a regular laptop. The code, datasets, and trained models are provided at \url{https://github.com/jbesty/residual_power_flow}.
\begin{figure}[ht]
    \centering
    \includegraphics[width=0.6\linewidth]{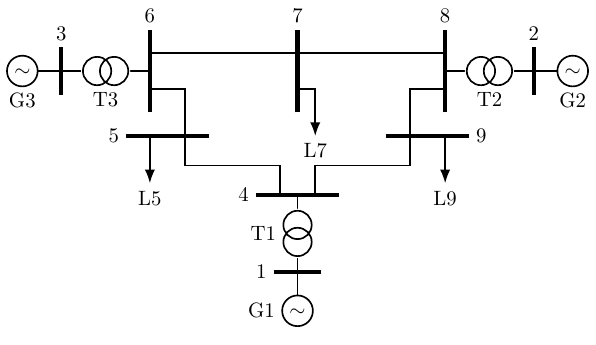}
    \caption{Single line diagram of the IEEE 9-bus system.}
    \label{fig:ieee_9bus_schematic}
\end{figure}
\subsubsection{Power system model}
We study the IEEE 9-bus system displayed in \cref{fig:ieee_9bus_schematic}. 
The model parameters stem from the MATPOWER case \cite{zimmerman_matpower_2011}. The following describes the modelling of the current injections $i_k$ in \cref{eq:static_injector,eq:dynamic_injector_SS}. We follow the convention that current flows \say{into} a bus have a positive sign.

All loads are modelled as constant power loads
\begin{align}
    i = h_{\text{load}}\inbrackets{\voltage{}, \controlis{k}} = -\frac{P_k}{V_k} - j \frac{Q_k}{V_k}
\end{align}
with $\controlis{k} = [P_k,  Q_k]$ and the voltage magnitude $V_k$ at the terminal bus. The current injections of the generators follow
\begin{align}
    i = h_{\text{gen}}\inbrackets{\voltage{}, \controlis{k}} = \frac{P_{M, k}}{V_k} + j \, K_{V, k} \inbrackets{ V_k - V_{ref,k}}
\end{align}
which represents a constant active power injection and a linear relationship between the reactive current injection and the voltage difference $V_k - V_{ref,k}$. The parameter $K_{V,k}$ governs the relation's strength and a high value of $K_{V,k}$ indicates stronger voltage support. We set $K_{V,k} = [130, 21, 13]$ for the three generators to approximate the characteristics of the models in \cite{sauer_power_1998}. The control variables are $\controlis{k} = [P_{M,k}, V_{ref, k}]$.

The current injections of the branches at the \say{from} and \say{to} terminal bus, indexed by $f$ and $t$, stem from a $\Pi$-model
\begin{alignat}{2}
    i = h_{\text{branch}, f} &= -y_{ff} V_f &&- y_{ft} V_t e^{j \branchAnglei{}}\\
    i = h_{\text{branch}, t} &= -y_{tf} V_f e^{-j \branchAnglei{}} &&- y_{tt} V_t
\end{alignat}
in which the branch angle \branchAnglei{} is positive in the direction $f \rightarrow t$. The branch admittances $y_{ff}, y_{ft}, y_{tf}, y_{tt}$ are defined as in \cite{zimmerman_matpower_2011}.

\subsubsection{Dataset generation}
The training and test datasets consist of 2000 and 1000 \glspl{OC} respectively. We generate the \glspl{OC} by sampling a total apparent power $S \in [1.0, 4.0]\si{\pu}$, and load shares $\eta_k$ to distribute $S$ randomly on the loads. For each load, we sample a power factor $\psi_k \in [0.9, 1.0]$ to obtain $P_{k} =  S \eta_k \psi_k $ and $Q_k =  S \eta_k \sqrt{1 - \psi_k^2}$. Similarly, we distribute $S$ across the active power setpoints $P_{M,k}$ of the generators by sampling participation shares $\eta_{M,k}$ and evaluating $P_{M,k} = S \eta_{M,k}$. The voltage setpoints are sampled independently from $V_{ref, k} \in [1.0, 1.05]\,\si{\pu}$. 

To generate a dataset with AC-feasible setpoints, that is $\residualScalarofVgivenU = 0$, we solve the \gls{RPF} in \cref{eq:power_flow_solution} while designating $P_{M,k}$ of one generator as the slack variable $u_s$. To generate a dataset with non-AC-feasible setpoints, we do not assign a slack variable and solve \cref{eq:rpf_unconstrained}. To increase the variation of infeasible points we alter $S$ for the generators by a factor between 1.0 and 1.08 which emulates the anticipation of grid losses.

\subsubsection{Neural solver training} We test two variants of neural solvers: first, linear features as in \cref{eq:neural_solver_linear}, and second, features learned with a \gls{NN} with two layers of width 100 and $\tanh{}$ activation functions as in \cref{eq:neural_solver_learned}. We train the \gls{NN} up to 6000 epochs using the L-BFGS algorithm \cite{liu_limited_1989}.

For the comparison with neural solvers based on the \gls{BIM}, we need to adjust the input features and prediction targets. This adjustment depends on the bus types: For PQ-buses, we use the active and reactive power injections $P_k$, $Q_k$ as inputs; for PV-buses (bus 2 and 3), we use the active power injection $P_k$ and voltage magnitude $V_k$; and for the slack bus (bus 1) the reference angle $\theta_1$ and the voltage magnitude $V_1$. The prediction targets include the bus angles $\theta_k$ at PV- and PQ-buses and voltage magnitudes $V_k$ at PQ-buses. After the prediction, $P_1, Q_1, Q_2, Q_3$ are calculated to match the corresponding constraints.


\subsection{Simpler predictions: Branch angles instead of bus angles}
\begin{figure}[bh]
  \centering
  \subfloat[Transformer T1 at reference bus 1]{%
    \includegraphics[width=0.95\linewidth]{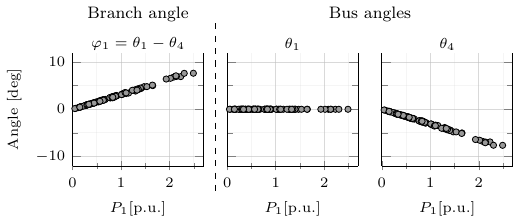}\label{subfig:angle_comparison_1}}
  \hfill
  \subfloat[Transformer T2 at bus 2]{%
    \includegraphics[width=0.95\linewidth]{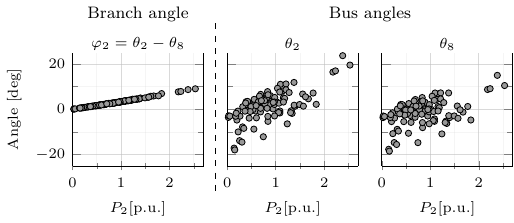}\label{subfig:angle_comparison_2}}
  \caption{Comparison of the angle variables for \gls{RPF} and the \gls{BIM}.}
  \label{fig:angle_comparison}
\end{figure}

We begin by a simple comparison of the distribution of the target variables \voltage{}. While the voltage magnitude $\voltageMagnitudei{}$ is unaffected by switching from a \gls{BIM} to \gls{RPF}, the branch angles \branchAngle{} differ from the common bus angle representation, denoted by $\theta$. We plot the angle variables of the transformers T1 and T2 in \cref{fig:angle_comparison} for the same set of \glspl{OC}. 
The x-axis shows the power setpoint $P_1$ and $P_2$ of the respective generator. According to the physics, the the branch angle $\branchAnglei{}$ increases nearly linearly with the power injection $P$. Using the branch angles as in the \gls{RPF} formulation, the expected linear function is clearly visible for both transformers. In contrast, the bus angles in the \gls{BIM} formulation show different patterns. The need to define a reference bus causes this obscuring of a simple linear relation. Furthermore, the larger the electrical distance of a branch to the reference bus, the larger this artefact. Hence, the artefact's strength varies across the grid.
\begin{figure}[!th]
  \centering
  \subfloat[Voltage magnitude predictions]{%
    \includegraphics[width=0.95\linewidth]{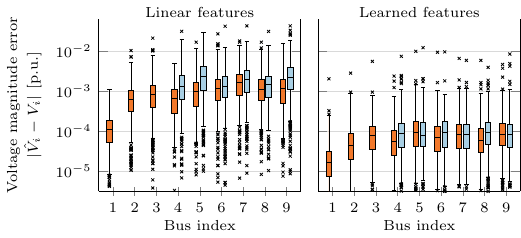}\label{subfig:ieee9_magnitudes}}
  \hfill
  \subfloat[Branch angle predictions]{%
    \includegraphics[width=0.95\linewidth]{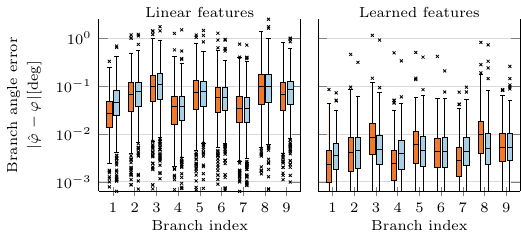}\label{subfig:ieee9_branch_angles}}
  \caption{Comparison of the prediction between the proposed \gls{RPF} formulation (\textcolor{RPF}{orange}) and the \gls{BIM} formulation (\textcolor{BIM}{blue}). The filled part of the boxplots represent the range of the 25th to 75th percentile, the whiskers indicate the 1.5-fold of the inter-quartile range and all points beyond are considered outliers represented as crosses.}
  \label{fig:ieee9_predictions}
\end{figure}
The \gls{RPF} formulation eliminates these representation asymmetries by using branch angles, which then simplifies their prediction.

\subsection{Prediction performance}
\Cref{fig:ieee9_predictions} compares the prediction performance of the \gls{BIM} and \gls{RPF} formulation for the voltages \voltage{}. The left and right panel in each subplot correspond to neural solvers with linear and learned features.
The learned features (right panels) increase the representation capacity of the neural solver, and, as expected, lead to lower errors; in this case roughly by a factor of 10. The performance of the \gls{RPF} and \gls{BIM} formulation is comparable as the features can compensate for representational asymmetries. The formulation change is more clearly visible for the linear features (left panels). The voltage magnitude prediction in \cref{subfig:ieee9_magnitudes} of the \gls{RPF}-based neural solver in orange consistently outperforms the \gls{BIM}-based one. The error for the voltage magnitudes at PV buses \voltageMagnitudei{1}, \voltageMagnitudei{2}, and \voltageMagnitudei{3} equals 0 for the \gls{BIM} formulation as they are supplied as input features. The voltage magnitudes \voltageMagnitudei{4}, \voltageMagnitudei{6}, and \voltageMagnitudei{9} show the largest difference between the formulations. We attribute \gls{RPF}'s improvement to the better representation of the strong voltage support at bus 1 which then impacts buses 4, 6, and 9. For the branch angles in \cref{subfig:ieee9_branch_angles} the performance differences are much smaller.

\subsection{Error distribution of residuals}
\begin{figure}[!th]
    \centering
    \includegraphics[]{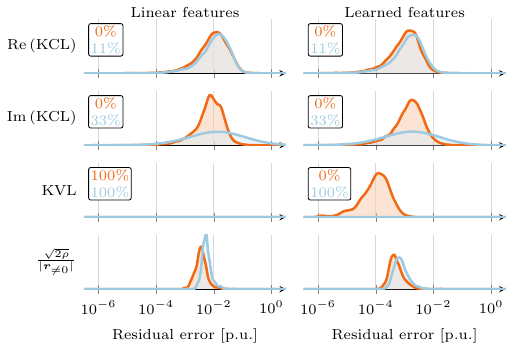}
    \caption{Distribution of residuals \residual{} across \glspl{OC} grouped by type for linear and learned features under the \gls{BIM} (\textcolor{BIM}{blue}) and \gls{RPF} (\textcolor{RPF}{orange}) formulation. The numbers in the top left corner indicate the share of residual values at numerical tolerance.}
    \label{fig:ieee9_residuals}
\end{figure}

While the analysis of the voltage prediction \voltageHat{} gives a first indication on the benefits of using \gls{RPF}, we now evaluate the residuals $\residual{}\inbrackets{\voltageHat; \control}$ based on the predictions, as they indicate the overall accuracy of the \gls{PF}. The error distributions in \cref{fig:ieee9_residuals} show that the performance varies across the type of residual shown as rows.
While the \gls{BIM} and \gls{RPF} formulation lead to similar error distributions for the real part of the nodal balance ($\KCL$), the imaginary part shows significantly higher error for the \gls{BIM} formulation. At the same time, 33\% of the residuals have no error, indicated in the box in the corner. These residuals occur at the PV buses as the reactive powers $Q_k$ are chosen after the prediction of \voltage{}. Thereby, the current balance will be satisfied, but the calculated values $\hat{Q}_k$ carry errors which affects line flows and hence neighbouring buses, leading to overall higher errors of the \gls{BIM} formulation. A similar phenomenon occurs for the slack bus power, but since only one bus is affected, the overall impact is less pronounced. The cycle balance $\KVL{}$ is satisfied by design for the \gls{BIM} formulation. The linear features can easily fit the branch angles to match the $\KVL{}$, whereas the learned features cause some error, but much smaller compared to the $\KCL$ errors. To judge the prediction performance overall, the residual \residualScalarofVHatgivenU{} serves as a good metric as it summarises the predictive performance in one number per \gls{OC}. In the last row of \cref{fig:ieee9_residuals}, we show the distribution of the average residual\footnote{We normalise by the number of non-zero residuals. In the \gls{BIM} formulation, $\imagOf{\KCL}$ residuals at PV buses, the $\KCL$ residuals at the slack bus, and $\KVL$ residuals are $0$ by design.}.
Overall, the choice of the \gls{RPF} formulation results in lower errors and more favourable error distributions. The residual norm \residualScalar{} of the \gls{RPF} forms a well-defined and simple metric to assess performance while learning. In contrast, the \gls{BIM} formulation introduces many artifacts that need to be considered when assessing the predictive performance.

\subsection{Learning from infeasible OCs}\label{subsec:learning_infeasible_OCs}
\begin{figure}
    \centering
    \includegraphics[width=0.95\linewidth]{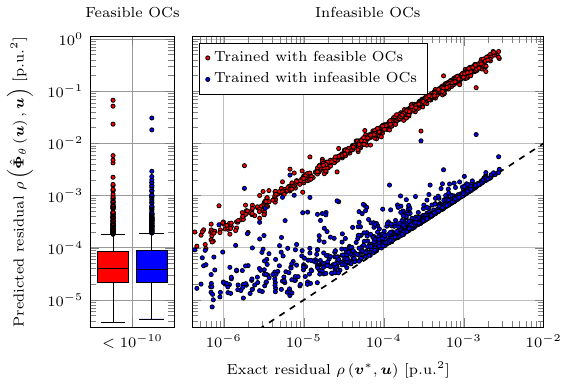}
    \caption{Prediction errors of feasible OCs (boxplots) and infeasible OCs (scatter plot). The colour indicates if the training was based on feasible or infeasible OCs. The dashed line indicates a perfect prediction.}
    \label{fig:prediction_error_infeasible_OCs}
\end{figure}

The \gls{BIM} formulation has no clear notion of a size of infeasibility as it is a binary characteristic of an \gls{OC}. In contrast, the \gls{RPF} associate infeasible \glspl{OC} with a residual value $\residualScalarofVgivenU{} > 0$ and defines the map $\voltage{}^*=\powerFlowVUMapOf{\control}$. With these definitions, we can meaningfully use infeasible \glspl{OC}, that is control inputs \control{} with $\residualScalarofVgivenU{} > 0$, in the training process of a neural solver \neuralSolverMapOfControl{}. \Cref{fig:prediction_error_infeasible_OCs} presents the comparison of training a neural solver only with feasible \glspl{OC} (in red) or with infeasible \glspl{OC} (in blue). When tested on feasible \glspl{OC}, the performance is nearly identical as shown in the boxplots. However, tested on infeasible \glspl{OC}, the residual error is more than ten times lower when using infeasible \glspl{OC} in the training. As the \gls{PO} approach entails optimising \residualScalar{}, predicting infeasible \glspl{OC} accurately becomes a crucial factor to success.

\section{Case study: Predict-then-Optimise with Neural Solvers for RPF}\label{sec:experiments}

We apply the neural solvers \neuralSolverMap{} with learned features to different \gls{PO} settings. We always use exactly the same neural solver as it was trained in \cref{sec:learning_problem}; no additional training or adjustments are required to the neural solver. All variations are achieved by altering the main optimisation problem.

\subsection{PO for AC-feasible Power Flow}
\begin{figure}[ht]
    \centering
    \includegraphics[]{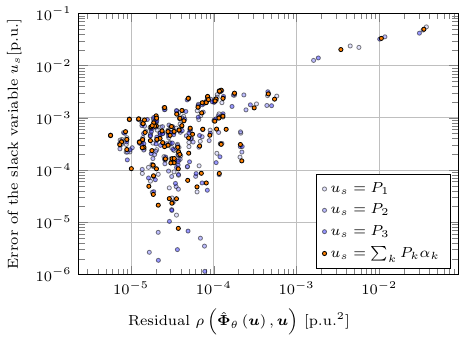}
    \caption{Error distribution for different slack variables $u_s$ (single generator and distributed) over the residual of the prediction.}
    \label{fig:PO_feasibility}
\end{figure}

We first solve for AC-feasible \glspl{PF}. The initial control \control{} of each \gls{OC} only meets the loss-less active power balance. The active power setpoint $P_{M}$ of one of the three generators or jointly in a distributed fashion forms the slack variable $u_s$. \Cref{fig:PO_feasibility} clearly shows that the necessary adjustment of the slack variable (on the y-axis) to achieve AC-feasibility has mostly errors less than \SI{0.001}{\pu}.
For \glspl{OC} with larger errors, the residual of the prediction indicates the lower accuracy. Thereby, the outliers in the top-right corner are easily spotted and can be traced back to higher errors of the neural solver.
Moreover, the solution accuracy is not only high for the slack variable but also for the voltage variables as they only rely on the accuracy of the neural solver as it was shown in \cref{fig:ieee9_predictions}.

\begin{figure}[ht]
    \centering
    \includegraphics[width=1.0\linewidth]{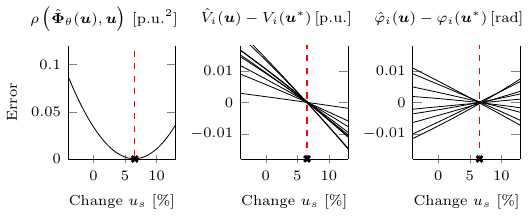}
  \caption{Relation between the residual value $\residualScalar{}\inbrackets{\neuralSolverMapOfControl, \control}$ evaluated with the neural solver and control values for which one element is adjusted as a slack variable. The minimum value of this curve corresponds to the solution $\hat{\control{}}^{\star} $ of the PO algorithm (black mark). The exact solution $\residualScalar{}\inbrackets{\voltage{}(\control{}^{\star}),\control{}^{\star}}$ is shown as the red dashed line.}
  \label{fig:PTO_slack_single_OC}
\end{figure}

The robustness of these results arises from the well-behavedness of the residual \residualScalar{} around feasible points. By starting from a loss-less power balance, we are often already close to a feasible solution and the gradients of \residualScalar{} are pointing towards feasibility. We illustrate this relationship for a single case in \cref{fig:PTO_slack_single_OC}. The value at 0\% change corresponds to the initial control with a low residual value $\residualScalar{}\inbrackets{\neuralSolverMapOfControl, \control}$ and also moderate voltage errors.

\subsection{PO for Quasi-Steady State Power Flow}
The \gls{PO} approach can be used for solving quasi-steady state \glspl{PF}. We introduce a droop characteristic for each generator
\begin{align}
    \Delta P_{M,k} = -\frac{1}{R}\frac{\omega - \omega_0}{\omega_0} P_{rated, k}
\end{align}
where the power adjustment $\Delta P_{M,k}$ is proportional to the rated power $P_{rated,k}$, the deviation of the frequency $\omega$ from the nominal value $\omega_0$ and the speed droop $R$, which we set to \num{0.04} \cite{cutsem_voltage_1998}. 
\begin{figure}[ht]
    \centering
    \includegraphics[width=0.8\linewidth]{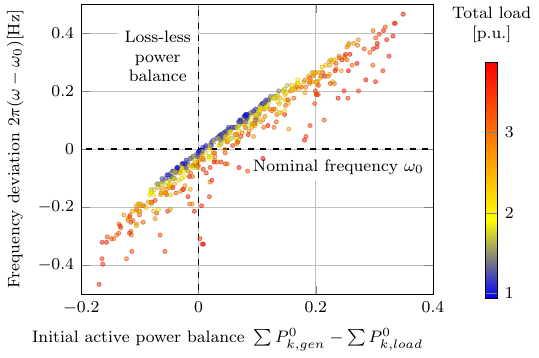}
    \caption{500 \glspl{OC} of different loading levels in quasi-steady state \gls{PF} solved with \gls{PO}. The initial setpoints $\control^0$ are adjusted by the frequency deviation.}
    \label{fig:PO_QSS_PF}
\end{figure}
The frequency $\omega$ becomes the slack variable $u_s$ that affects the initial control inputs $\control{}^0$. We sample 500 infeasible \glspl{OC} and report the frequency deviations in \cref{fig:PO_QSS_PF}. This result shows the versatility of the \gls{RPF}-based \gls{PO} approach: We never trained the neural solver for quasi-steady state \gls{PF}, we only changed the slack variable $u_s$ to solve this power system task.

\subsection{PO for AC-Optimal Power Flow}
\begin{figure}[ht]
  \centering
  \subfloat[Residual contours.]{%
    \includegraphics[width=0.95\linewidth]{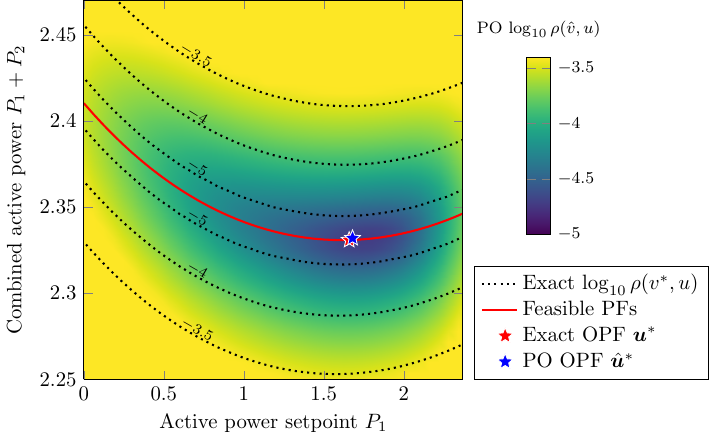}\label{subfig:PO_AC_OPF_residual}}
    \hfill
  \subfloat[Cost function contours.]{%
    \includegraphics[height=3.3cm]{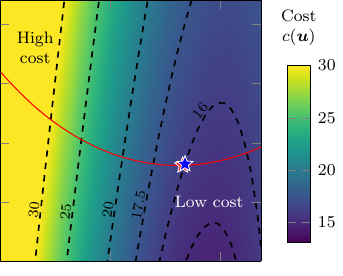}\label{subfig:PO_AC_OPF_cost}}
  \hspace{0.1cm}
  \subfloat[Operational constraint contours.]{%
    \includegraphics[height=3.3cm]{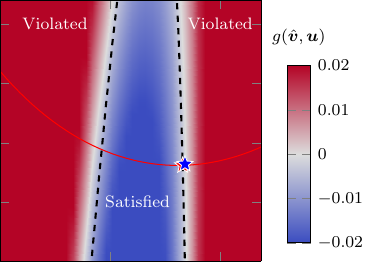}\label{subfig:PO_AC_OPF_constraint}}
  \caption{Contours of the terms in the \gls{PO} problem  based on the neural solver (coloured) and when exactly solving the \gls{RPF} in \cref{eq:rpf_unconstrained} (lines).}
  \label{fig:PO_AC_OPF}
\end{figure}

Lastly, we apply \gls{PO} to an AC-\gls{OPF} problem to showcase the seamless integration of a \gls{RPF}-based neural solver. Based on the neural solver prediction \voltageHat{}, we evaluate the residual \residualScalar{} and the satisfaction of the operational constraints. For variations of two generator setpoints $P_1$ and $P_2$, \cref{subfig:PO_AC_OPF_residual} shows the resulting contours of the residual and \cref{subfig:PO_AC_OPF_constraint} the constraint satisfaction. 
By overlaying these contours with the cost contours in \cref{subfig:PO_AC_OPF_cost}, we can identify the optimal control values \controlstarHat{} based on \gls{PO}, marked by the blue star. The result is close to the exact optimal value \controlstar{} (red star). The difference between \controlstarHat{} and \controlstar{} stems from the approximation quality of the neural solver. The more accurate the solver approximates the \gls{RPF} solution, the better the predicted contours (coloured) in \cref{fig:PO_AC_OPF} match the exact ones shown as dashed lines. For higher-dimensional problems, the solution of \cref{eq:PO_AC_OPF} can be performed using gradient-based methods as all evaluations are differentiable. 

\section{Discussion}\label{sec:discussion}

The \gls{RPF} formulation opens many paths forward around neural solvers and the \gls{PO} approach, but also in connection with established \gls{PF} topics. 

\subsubsection*{Advancing Neural Solvers for RPF}
With the proposed \gls{RPF} formulation, we see the need to revisit neural solver architectures and training. In particular, the avoidance of bus types could allow architectures that align closely with \gls{RPF}; graph-based neural solvers as in \cite{donon_neural_2020} seem promising. To improve the training additional physics-informed regularisers could be added. The optimisation of the training and evaluation computations will be an important focus for scaling neural solvers. For reference, the used non-optimised implementation in Julia trains the neural solver within a few minutes and evaluates the \gls{RPF} in about \SI{30}{\micro\second}. By tailoring and optimising the computations of the residual calculations, the time and memory footprint is likely to decrease.

The ambition for neural solvers should to be to solve a wide range of \glspl{OC}, ideally with topological changes, component availabilities, and varying parameter settings. This adaptability would enable the use of the trained neural solvers for a wide variety of \gls{PO} formulations. Such a foundational \gls{RPF} neural solver would even justify high upfront cost for dataset generation and training due its versatile applicability. 

\subsubsection*{Tailoring PO to Neural Solvers}
The robustness, speed, and accuracy of the \gls{PO} approach hinges around the setup of the outer optimisation problem in \cref{eq:PO_formulation_generic}. The applied optimisation algorithms need to be aligned with the constraint definition and the weighting of the terms in the objective function. Furthermore, different neural solver architectures could exhibit different characteristics when included in the \gls{PO} approach. 

We have only showcased a few power system task that suit the approach of \gls{PO} with neural solver for \gls{RPF}. State estimation appears like a natural fit and more advanced neural solver architectures, as discussed above, could enable contingency analyses and planning studies. These application would require to handle topological adjustment within the neural solver. Furthermore, dynamics studies could be conducted if dynamic current injections in \cref{eq:dynamic_injector} can be considered.

\subsubsection*{New perspectives through RPF}
Our development of \gls{RPF} was driven by its use for neural solvers. However, we observed many connections to classical power system topics for which \gls{RPF} and its quantification of infeasibility could yield new perspectives. To name a few, state estimation and dynamic simulations are tightly linked, but also methods like continuation \gls{PF}, holomorphic embeddings of \gls{PF}, \gls{PF} relaxations, and the study of feasibility of \glspl{OC} seem relatable. 

\section{Conclusion}\label{sec:conclusion}
\glsresetall
We proposed the \gls{RPF} formulation that is tailored for learning power flows with neural solvers. \gls{RPF} has two key characteristics that make it particular suitable for learning: First, the formulation avoids the asymmetries that are usually introduced by defining bus types. Second, infeasible power flows are naturally represented which improves the handling of approximation errors. These characteristics of \gls{RPF} improve the suitability for learning power flows with neural solvers while also allowing their direct integration into a \gls{PO} approach. While the \gls{PO} approach provides flexibility to adapt to many power flow-related tasks, the same neural solver can be reused to provide fast power flow approximations.

This work aimed at providing a solid basis to develop neural solvers into a practical solution approach for power flow constrained problems. The proposed \gls{RPF} formulation allows focusing separately on two directions for improvements: First, improving the accuracy and training of neural solvers for \gls{RPF}, and second, the efficient integration in the \gls{PO} setting to cover different power system tasks. 

\appendices



\bibliographystyle{IEEEtran}
\bibliography{bstcontrol.bib,references.bib}

\ifCLASSOPTIONcaptionsoff
  \newpage
\fi






\end{document}